\documentclass[prl,showpacs,aps,twocolumn,nofootinbib,superscriptaddress]{revtex4}% Physical Review B
\usepackage{graphicx,amssymb,amsmath}

\begin{document}

\title{Evidence of 1D behaviour of He$^4$ confined 
within carbon-nanotube bundles}

\author{J. C. Lasjaunias}

\affiliation{Centre de Recherches sur les Tr\`{e}s Basses Temp\'{e}ratures, 
CNRS, BP 166, 38042 Grenoble cedex 9, France}

\author{K. Biljakovi\'{c}}

\affiliation{Institute of Physics, Hr-10 001 Zagreb, P.O.B. 304, Croatia}

\author{J. L. Sauvajol}

\affiliation{Groupe de Dynamique des Phases Condens\'{e}es, Universit\'{e} 
Montpellier II, 34095 Montpellier Cedex 5, France}

\author{P. Monceau}

\affiliation{Centre de Recherches sur les Tr\`{e}s Basses Temp\'{e}ratures, 
CNRS, BP 166, 38042 Grenoble cedex 9, France}
\affiliation{Laboratoire Leon Brillouin, CEA-CNRS, CEA Saclay 91191 Gif-sur-Yvette cedex, France}

\begin{abstract}
We present the first low-temperature thermodynamic investigation of the
controlled physisorption of He$^{4}$ gas in carbon single-wall
nanotube (SWNT) samples. The vibrational specific heat measured between 100
mK and 6 K demonstrates an extreme sensitivity to outgassing conditions. 
For bundles with a few number of NTs the extra contribution to the specific heat, C$_{ads}$,
originating from adsorbed He$^{4}$ at very low density displays 1D behavior, 
 typical for He atoms localized within linear channels as grooves and 
interstitials, for the first time evidenced. For larger bundles, C$_{ads}$
recovers the 2D behaviour akin to the case of He$^{4}$ films on planar
substrates (grafoil).
\end{abstract}

\date{20/1/ 03}
\maketitle

The discovery of carbon nanotubes is very exciting from the academic point
of view, but also for engineering purposes. Reducing dimensionality can bring
entirely new properties for the thermodynamics of vibrational states for
nanotubes as well as for the mechanisms of gas adsorption (in structures
like bundles), in particular He$^{4}$, which has already been largely
investigated in the 2D case. 1D confinement of vibrational modes was
supposed to explain the specific heat (C$_{p}$) in SWNT measured down to 2 K 
\cite{1}. He within the interstitials in nanotube bundles was expected to
condense into a lattice gas \cite{2}. Our recent C$_{p}$ measurements did not
support these both expectations \cite{3}; but they yielded C$_{p}$ results 
of pristine nanotubes and showed a remarkable influence of a relatively
small quantity of adsorbed He, which in particular allows the comparison to
the case of He films in 2D configurations, already largely investigated \cite
{4,5,6}.

In this Letter we present the low-T specific heat investigation on two SWNT
samples in similar conditions as far as adsorbed He$^{4}$ is concerned. The
first sample was prepared by laser vaporization (sample LV) and the second
one by electric-arc discharge (sample AD). X-ray and neutron diffraction,
and Raman experiments were systematically used for the characterization of
both samples \cite{7,8}. 

{\underline{Sample LV}} (synthesized in ``Rice University conditions'' \cite
{9}) was obtained by laser vaporization at 1100$^{\circ }$C with Ni and Co
catalysts, up to at. 2\% as residual concentration in final samples. The
SWNTs of mean diameter 1.4 nm, with a standard deviation of 0.2 nm \cite{8}
form bundles organized in the usual hexagonal lattice of parameter 1.7 nm.
Bundles with a diameter of 10-13 nm contain typically 30-50 tubes, with an
extension up to 20 nm (i.e. up to N$_{t}\sim $100 tubes). The LV sample (m=45 mg) in
form of several elastic foils (buckypaper as in ref. \cite{1}) was used for
previous heat capacity measurement \cite{3}. {\underline{Sample AD}} was
synthesized at Montpellier by arc-discharge \cite{10} with Ni (0.5\%) and Y
catalysts (0.5 \%), in total 1 at. \%, and with no Co. The individual SWNT
of similar diameter as in sample LV \cite{7,8} here are organized in bundles
generally smaller in diameter -- up to 10 nm with N$_{t}$ less than 20-30 tubes. The
AD sample (m=90 mg) was in a form of pellet (13 mm in diameter, 1 mm thick)
obtained by pressurisation of powder with 10 kbar \cite{11}. Within the
usual purification technique, the nanotubes are close-ended. So, we suppose
that no adsorption can occur within the tubes, and only interstitials or
external grooves and surfaces of the bundles are the most probable occupancy
sites. Both samples were mounted in the same manner \cite{3} in a glove box
under nitrogen atmosphere to avoid at maximum gaseous contamination from
ambient air and thereafter transferred in the dilution cryostat. The data
were collected by the transient heat pulse method between 100 mK and 6 K.

We have studied the adsorption in LV and AD samples by simply using our
usual cooling procedure down to 4.2 K, He$^{4}$ being the exchange gas \cite
{3}. After initial pumping at room temperature, He$^{4}$ is introduced in
the calorimeter chamber at 77 K at a pressure of 6-7 10$^{-2}$ mbar.
Normally, when the sample does not adsorb, at 4.2 K we pump out the exchange
gas in order to get good conditions for C$_{p}$ measurements ($\sim $10$^{-6}$
mbar). Spectacularly, in LV samples all the He$^{4}$ was adsorbed, so that
cooling down to He temperature was not possible without a new injection!
This was the first evidence of the strong He adsorption in SWNT samples. As
it is well established that He$^{4}$ is totally desorbed by reheating up to
25-30 K \cite{1,3,12}, we used this procedure to obtain the first desorbed
state - run A \cite{3}. In addition, a further long secondary
pumping (a few days to a whole week) at 300 K, coupled with the above very
efficient desorbing procedure (reheating from 4.2 to 40 K) was used to
define the heat capacity in the best outgassing conditions (run C for LV
sample \cite{3} and III in AD sample).

\begin{figure}[tbp]
\centerline{\includegraphics[width=8cm]{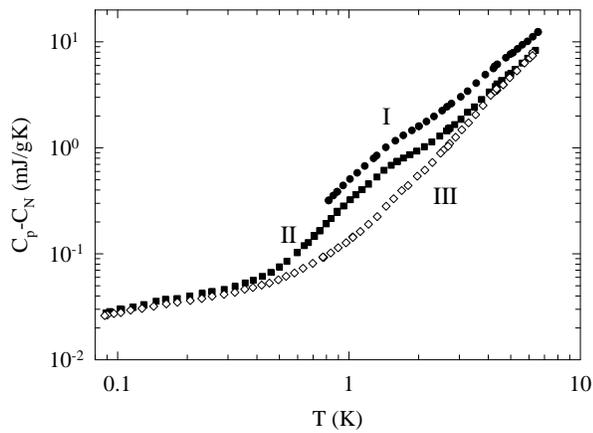}}
\caption{Vibrational specific heat C$_{vib}$=C$_{p}$-C$_{N}$ of the AD
sample (per g of sample): pristine state (without He) - III; intermediate adsorption - II;
maximum adsorption in our experimental conditions (see text) - I.}
\label{fig1}
\end{figure}

Figure~\ref{fig1} presents the new data collected on the AD sample. {%
\underline{State I}} was obtained with the usual He dose for cooling the
sample down to 4.2 K and successive pumping under secondary vacuum at 5 K. {%
\underline{State II}} corresponds to more efficient outgassing conditions -
reheating under active pumping up to 15 K. As this temperature is too low
for a complete desorption, this run represents a case of intermediate He
adsorption. The most efficient desorption procedure was performed for {%
\underline{State III}} (equivalent of run C for LV in Ref. \cite{3}). First
the sample was under active secondary pumping during one week at 300 K.
After it was cooled down to 4.2 K (following procedure I). Finally, further
reheating to 35-40 K enabled complete outgassing of He. Consequently, we
take C$_{p}$ of {\underline{State III}}  as the \textit{pristine} nanotube value,
corresponding to the vibrational contribution. Here we only indicate that
the extrinsic contribution of the ferromagnetic catalyst, which shows up via
a nuclear hyperfine term C$_{N}$ (rising below 0.15 K), was much smaller
than in LV sample due to the absence of Co \cite{13}. This term is
subtracted and the {\underline {vibrational}} heat capacity C$_{vib}$= C$_{p}$ - C$_{N}$ is
reported for the three runs in Figure~\ref{fig1}. There is an evident and
progressive decrease of C$_{p}$ on improving the outgassing. In comparison
to the pristine data (State III), adsorption of He results in a broad anomaly
for state II entirely included in our investigated T-range. This was also
the case in the LV sample (run A compared to C in ref. \cite{3}).

Another contribution becomes predominant below 0.5 K, which is \textit{not
related} to He adsorption, as it is almost the same for both states II and
III \cite{14}. It exists in the same T-range also in the LV sample and it
can be represented by a sub-linear power law $\sim $ T$^{\alpha }$, with $%
\alpha $=0.34 in AD and $\alpha $=0.62 in LV. This contribution is very
probably due to localized low energy excitations (defects in the carbon
lattice and/or the nanotube itself). Finally, C$_{vib}$ in the pristine
state (III) of the AD sample is the sum of two terms for T$\leq $1.8 K: C$%
_{vib}$= AT$^{\alpha }$ + $\beta $T$^{3}$ = 0.061 T$^{0.34}$+ 0.072 T$^{3}$
(mJ/gK). The second term represents the Debye phononic contribution, which
indicates the 3D character of the bundles for low-frequency phonon modes.
The cubic regime is no more obeyed above 1.8 K, where C$_{vib}$ changes
smoothly to a quadratic regime. The lower crossover temperature T$_{co}$ in
AD indicates that the transversal lattice cohesion is weakened compared to
LV \cite{3}, which is consistent with a smaller number of tubes N$_{t}$ in
the bundle. This softening of the inter-tube cohesive forces - decrease of transverse sound velocity - 
is also in agreement with the larger $\beta $: being 0.072 mJ/gK$^{4}$ in AD compared
to 0.035 mJ/gK$^{4}$ in LV \cite{3,15}.

\begin{figure}[tbp]
\centerline{\includegraphics[width=8cm]{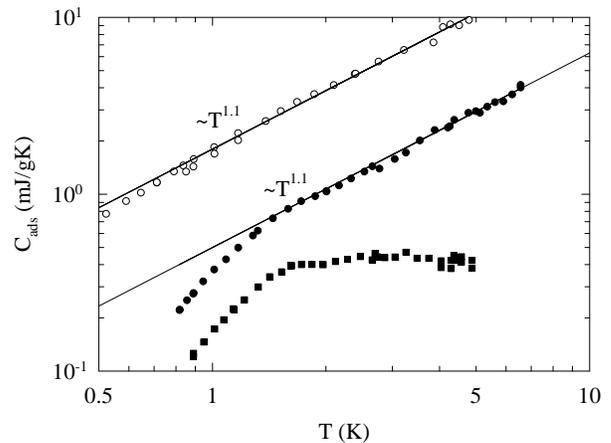}}
\caption{Specific heat of He$^{4}$ adsorbate (after subtraction of the
corresponding pristine values; C$_{ads}$ per g of sample), for LV and AD sample.
Data ($\circ$) and ($\bullet$) are for maximum adsorption for LV sample (3 at.\% of C
atoms) and AD sample (1.5 at.\% ), respectively. Data ($\blacksquare $)
correspond to the AD sample at intermediate adsorption.}
\label{fig2}
\end{figure}

\begin{figure}[tbp]
\centerline{\includegraphics[width=8cm]{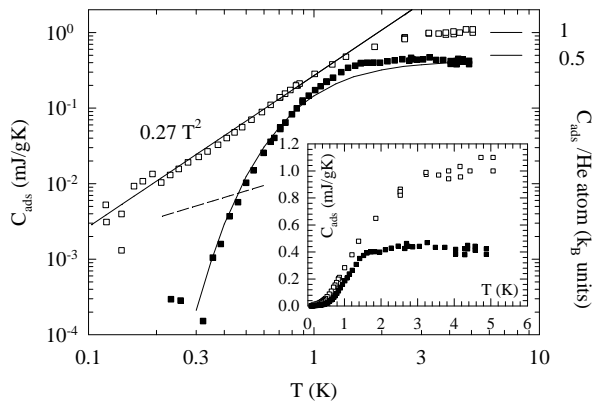}}
\caption{Specific heat of He$^{4}$ adsorbate at intermediate adsorption: ($\square $) state A of the LV
sample, ($\blacksquare $) state II of the AD sample. Data (%
$\blacksquare $) below 1 K are fitted by an Einstein specific heat, with $%
\Theta _{E}$=3.8 K, and data ($\square $) by a quadratic variation below 1.5
K. K. The same data are plotted in a linear plot in the inset.  The plateau
values recalculated per atom of He correspond to 1 k$_{B}$ for LV and 0.4 k$%
_{B}$ for AD sample. The dashed line is the residual term which was removed in the case of AD sample \cite{14}.}
\label{fig3}
\end{figure}

To go further with the analysis, we make the assumption that the
contribution of adatoms to the specific heat is additional to that of the
pristine SWNT. It holds to a great accuracy due to the fact that the adsorbate modes couple only weakly to the substrate
vibrational modes (the substrate atomic mass mismatch is large - 4/12).
Hence we obtain C$_{ads}$ after subtracting the corresponding pristine
values; run C for LV \cite{3} and state III for AD. C$_{ads}$ for the
maximum adsorption is presented in Figure \ref{fig2} and for the case of
intermediate adsorption in Figure \ref{fig3}. Here we repeat that the
pressure of He$^{4}$ gas admitted in the sample chamber was the same in both
cases (corresponding to N$_{0}$=7$\cdot$10$^{19}$ atoms). However the AD sample
having two times larger mass, the maximal number of accessible He atoms per
one C atom was 3\% in LV sample and 1.5\% in the AD sample; this fact
together with the different sample topologies (characterized with N$_{t}$)
will be crucial in our interpretation of C$_{ads}$.

C$_{ads}$ represents at least 50\% of the total C$_{p}$ (at 1 K) in our
conditions. It exhibits a \textit{striking similarity} for both samples
which certainly reflects the same thermodynamics of the additional contribution
coming from the adsorbed He. Figure \ref{fig2} shows that in the case of
maximum adsorption both samples follow a close to linear-in-T behaviour,
with a smooth crossover to a stronger T dependence below 1 K. However, if we
take into account that the number of available He atoms per one C atom is
two times larger for LV sample, the relative amplitude of C$_{ads}$ of LV is
still almost two times larger than for AD sample.

For the intermediate adsorption state (Figure \ref{fig3}) there
is again a similar qualitative behaviour for both systems; after an initial
increase, there is a saturation towards a plateau which extends from 1.5 to
5 K for AD or from 3 K to 5 K for LV sample. The value of C$_{ads}$ per He
atom in the plateau region should indicate the effective dimensionality for
the He atomic excitations, being 0.5 k$_{B}$ for a 1D gas or 1 k$_{B}$ for a
2D fluid (gas or liquid). The rise of the plateau over 1 k$_{B}$ corresponds
to the occupation of excited single-particle states in the direction
perpendicular to the surface, i.e. announcing the 3D ordering (as the
completion of the first 2D monolayer).

For the estimation of the corresponding number of He adatoms in the
partially outgassed states (Figure \ref{fig3}) for each sample, we have used
informations from the desorption experiments, which were performed in quite
similar conditions as here \cite{12,16}. For two outgassing conditions
corresponding to different desorption temperatures, T$_{D}$$\sim $15 K for
AD and 25-30 K for LV sample, the remaining numbers of He atoms were
estimated as N$_{AD}\sim $N$_{0}$/10 and N$_{LV}\sim $ N$_{0}$/20, i.e. N$%
_{AD}\sim $ 7 10$^{18}$ and N$_{LV}\sim $ 3.5 10$^{18}$ \cite{17}.
Consequently, using absolute values of C$_{ads}$ given in the inset of
Figure \ref{fig3} - 38 $\mu $J/K for AD (90 mg) and 45 $\mu $J/K for LV (45
mg) -- the corresponding plateau values per one He atom are 0.4 k$_{B}$ for
AD and 0.95-1.0 k$_{B}$ for LV.

It is very difficult to reconcile these extreme cases (2D-LV behavior and
1D-AD behavior) if one only considers the \textit{absolute number} of He
adatoms, N$_{AD}>$ N$_{LV}$, which seems to lead to the inverse effect! In
fact, contrary to the case of He films on planar substrates where the He
behavior is surprisingly independent of the structure of the substrate \cite
{5}, here we are in the case where the \textit{topology of the substrate
plays an essential role}. This appears to be a natural conclusion in the
case of C-nanotube bundles. That finally gives the opportunity to detect 1D
effects in confined He (in a very diluted limit), which was not possible (or
evidenced) in low-T He films.

The essential topological difference between AD and LV samples is the
main \textit{size of the bundles} \cite{7,8}. In order to make the most
simple analysis, we neglect the certain distribution in size and we suppose
that AD bundles have 19 tubes and LV 61 (corresponding to 3 or 5 tubes on
each facet in regular hexagonal arrays). There are three specific sites of
adsorption for He \cite{19}: intersticials (IC), outer grooves (G) and
bundle surface (S). Recent adsorption isotherm measurements \cite{20},
consistent with He desorption experiments \cite{12,16}, indicate that
adsorption is first in the outer grooves of the bundles, then filling the
rest of the outer surface, and finally layering on the outer surface (in
successive layers). Monte-Carlo simulations \cite{21} show an extreme
pressure sensitivity and an abrupt crossover from 2D coverages to 1D
filling, which successively leaves G and IC channels as the last refuge for
the He atoms on progressive lowering pressure. From the above N$_{AD}$ and N$%
_{LV}$ values, we now estimate, using these topological
considerations, the possible occupied places, either in outer 2D graphene
surfaces or 1D (IC and G) channels. 

For this purpose, we make the following assumptions: we start from an
effective adsorption surface of $\sim $300 m$^{2}$/g per nanotube, in good
correspondence with 80 m$^{2}$/g for a typical bundle of N$_{t}$=37 \cite{20}%
. Now we suppose that a complete coverage of external surfaces (2D sites)
varies between 12 and 9 He sites (between 2 grooves on facets) for AD and LV
(N$_{t}$=19 and 61 repectively). Taking also in account the possible
adsorption in the 1D channels (G + IC), this yields S$_{eff}$=100 m$^{2}$/g
for N$_{t}$=19, and 70 m$^{2}$/g for N$_{t}$=61, with 80\% and 66\% of sites
in surface, respectively. For our sample masses, we get 2.0 m$^{2}$ of
effective surfaces for 1D channels and 7.0 m$^{2}$ for external surface in
AD sample, and 1 m$^{2}$ for 1D and 2.0 m$^{2}$ for surface of LV sample. We
suppose that adsorption takes place in the same external conditions (i.e.
supposing similar binding potentials) and that 12 atoms/nm$^{2}$ is a
complete layer coverage, as for He-films. Consequently, the {\underline{%
final configurations}} for the total desorption of N$_{0}$ atoms of He at
T=4 K should be for {\underline{AD sample}}: all 1D channels occupied and
around 1/2 of the outer surface of bundles and for {\underline{LV sample}}:
all 1D channels, 2 external layers occupied, and 50\% of a third layer.

Outgassing starts from the outer surfaces and continues with atoms at
the lower surfaces. In the case of AD sample (T$_{D}$=15 K and remaining N$%
_{AD}\sim $7 10$^{18}$), first all He atoms of the surface ($\sim $5 10$^{19}
$) were degassed \cite{19,21}. Further degassing empties the groove regions 
\cite{22} and probably also some weakly bond IC sites (in total 0.7 10$^{19}$%
) \cite{21}. Finally, He remains only in some IC sites \cite{21,22}, i.e.
about 50\% of the IC sites are still occupied. This is the origin of 1D He
confinement behaviour in AD as demonstrated in Figure \ref{fig3}. Also
consistent with our 1D interpretation for the AD sample is the very rapid
decrease of the He heat capacity below 1K, which can be well fitted with an
Einstein function, varying like ($\Theta _{E}$/T)$^{2}$exp(-$\Theta _{E}$/T),
with $\Theta _{E}$=3.8 K (Figure \ref{fig3}). This indicates independent
harmonic oscillators, probably localized. At our knowledge, the exponential
decay has never been observed in He$^{4}$ on grafoil. It was supposed to
occur at very low coverage, i.e. with a very feeble C$_{p}$ signal \cite{4,5}%
.

For the LV sample, outgassing is of a higher cost in energy because one has
to outgass the successive layers and the corresponding desorption energy
should be much larger for completely removing the 1$^{st}$ layer \cite{20}.
The heat of adsorption of the second layer for He on grafoil is very large
(35 K) in comparison to the third (15 K) and the following \cite{4}. We
conclude that, even at a desorption temperature of 25-30 K, there remains
still a considerable number of 2D He coverages for this sample. This is
consistent with the 2D behaviour of C$_{ads}$, which clearly shows a
quadratic regime over more than one decade in T; C$_{ads}$= 0.27-0.28 T$^{2}$
mJ/g K  (Figure \ref{fig3}). This is signature of a 2D
--fluid or solid-regime for adsorbed He. For the fluid case, consistent with
the limit C/R$\rightarrow $1 (R is the gas constant), for the molar heat
capacity of He, the corresponding 2D Debye temperature for only longitudinal
phonons is: C$_{ph}$=14.42 R (T/$\theta $)$^{2}$\cite{23}, so that C$_{ads}$=2.1 T$^{2}$ J/molK yields $%
\theta $=7.5 K. This value can be compared to 5.1 K obtained in the same
T-range (0.1-0.5 K) for the first layer of He$^{4}$ on grafoil \cite{4}. In
addition, the amplitude of the T$^{2}$ regime (or the Debye T) is rather
independent of the order of the layer up to the third one \cite{4}.

In conclusion, we gave the first thermodynamical evidence of 1D confinement
of He in NT bundles, as well as 2D behavior of C$_{ads}$ for bundles with larger size. 
These results suggest further studies of the crossover
between 1D to 2D adsorbate behavior by varying, for a given topology of the NT bundles, 
the concentration of He atoms in the dilute limit. This is of great significance, since it may
give basic informations on the very low energy states of He, such as
zero-point energy, localization, etc /dots.  The remarkable absence of peaks of C$_{ads}$ in this T range
indicates a strong influence of size-effects imposed by NT bundles compared to
conventional He films, such as on grafoil. This could be accounted for by a much shorter 
correlation length which inhibits the developement of structural phase transitions. 
One result which remains to be explained is C$_{ads}\sim $T$^{1.1}$ for rather large adsorption levels (a few at.\%) 
of He, when the adsorbate behavior is essentially
determined by the external ''dressing'' of the bundles. This seems to be a
completely new physics in comparison to the previous results obtained for
the successive layering processes on purely 2D substrates like grafoil and
deserves some theoretical considerations. 

\begin{acknowledgments}
A. \v{S}iber, L. Firlej, B. Kuchta and H. Godfrin are grately
acknowledged for stimulating discussions and D. Stare\v{s}ini\'{c} for his help.
\end{acknowledgments}


\begin{thebibliography}{99}
\bibitem{11}  This pressurisation results in a preferential organization of
the bundles parallel to the flat surfaces of the pellet and possibly a
deformation of cylindrical shape of NTs. Similarly, in LV samples,
filtration/aspiration process may result also in a preferential orientation
of the bundles parallel to the foil surfaces.
\bibitem{13}  C$_{N}$=0.15 T$^{-2}$ (\ensuremath{\mu}J/gK) found for AD is
25 times smaller in amplitude than for LV \cite{6}. This strong reduction is
mainly a consequence of the much smaller effective magnetic field H$_{eff}$
in Ni compared to Co, responsible for the splitting of nuclear energy levels.
\bibitem{14}  A slight variation (7\%) of the power law T$^{0.34}$
contribution between run II and III of AD sample (see Fig 1), was analysed
as 0.015 T$^{0.9}$ mJ/g K (dashed lines in Fig.3) and subtracted in order to
obtain the residual C$_{ads}$ (\underline{due only to adsorbed He})
\bibitem{17}  Despite uncertainty in the absolute values
(factor of two) our estimation yields a convincing self consistency with our
experimental data as shown in the following.
\end{thebibliography}
\end{document}